\documentclass[aps,prd,letterpaper,showpacs,twocolumn,superscriptaddress,nofootinbib,floatfix]{revtex4}

\usepackage{amssymb}
\usepackage{amsmath}
\usepackage{verbatim}
\usepackage{mathrsfs}
\usepackage{amsfonts}
\usepackage{latexsym}
\usepackage{epsfig}
\usepackage{color}
\usepackage{graphicx,subfigure}
\usepackage{units}
\usepackage{slashbox}

\begin{document}


\title{Cosmological scalar field perturbations can grow}

\author{Miguel Alcubierre}
\affiliation{Instituto de Ciencias Nucleares, Universidad Nacional
  Aut\'onoma de M\'exico, Circuito Exterior C.U., A.P. 70-543,
  M\'exico D.F. 04510, M\'exico}
  
\author{Axel de la Macorra}
\affiliation{Instituto de Fisica, Universidad Nacional
  Aut\'onoma de M\'exico, Circuito Exterior C.U., A.P. 20-364,
  M\'exico D.F. 04510, M\'exico}  

\author{Alberto Diez-Tejedor} \affiliation{Santa Cruz Institute for
  Particle Physics and Department of Physics, University of
  California, Santa Cruz, CA, 95064, USA} \affiliation{Departamento de
  F\'isica, Divisi\'on de Ciencias e Ingenier\'ias, Campus Le\'on,
  Universidad de Guanajuato, Le\'on 37150, M\'exico}

\author{Jos\'e M. Torres}
\affiliation{Instituto de Ciencias Nucleares, Universidad Nacional
  Aut\'onoma de M\'exico, Circuito Exterior C.U., A.P. 70-543,
  M\'exico D.F. 04510, M\'exico}


\date{\today}


\begin{abstract} 
It has been argued that the small perturbations to the homogeneous and
isotropic configurations of a canonical scalar field in an expanding
universe do not grow.  We show that this is not true in general, and
clarify the root of the misunderstanding.  We revisit a simple model
in which the zero-mode of a free scalar field oscillates with high
frequency around the minimum of the potential. Under this assumption
the linear perturbations grow like those in the standard cold dark
matter scenario, but with a Jeans length at the scale of the Compton
wavelength of the scalar particle.  Contrary to previous analyses in
the literature our results do not rely on time-averages and/or fluid
identifications, and instead we solve both analytically (in terms of a
well-defined series expansion) and numerically the linearized
Einstein-Klein-Gordon system.  Also, we use
gauge-invariant fields, which makes the physical analysis more
transparent and simplifies the comparison with previous works carried 
out in different gauges. As a byproduct of this study we identify a time-dependent
modulation of the different physical quantities associated to the
background as well as the perturbations with potential observational
consequences in dark matter models.
\end{abstract}




\maketitle


\section{Introduction}
\label{sec:intro}

It is difficult to overestimate the relevance of the Jeans instability
in modern physical cosmology.  In order to understand the emergence of
cosmic structure, we need a mechanism that transforms the nearly
homogeneous and isotropic early universe we infer from e.g. the cosmic
microwave background observations, to the highly clumped one we can
see today at the scale of galaxy clusters and below.  In the standard
cosmological scenario this transition is possible thanks to the
instability of a scalar mode that appears when we couple dark matter
(DM) to gravity, the {\it gravitational Jeans instability}.

In a universe dominated by a barotropic perfect fluid with equation of
state $p=p(\varepsilon)\ll\varepsilon$, the behavior of the small
perturbations in the energy density depends crucially on the speed of
sound, $c_s^2=\partial p/\partial\varepsilon$, and the Hubble radius,
$H^{-1}$ (throughout this paper we use natural units such that
$c=\hbar=1$).  Roughly speaking, we can distinguish three different
regions in Fourier space~\cite{Mukhanov}: $i)$~On scales smaller than
the Jeans length, $\lambda_J = c_s (\pi/G\varepsilon_0)^{1/2}\sim c_s
H^{-1}$, the contrast in the energy density oscillates with damping
amplitude, due to the stabilizing effect of pressure and the expansion
of the universe, respectively.  Here $G$ is the Newton constant and
$\varepsilon_0$ is the background matter density.  $ii)$~Above this
length scale but still below the Hubble radius, self-gravity dominates
and the contrast in the energy density grows: the Jeans instability
comes into play.  $iii)$~Finally, at scales larger than the Hubble
radius a relativistic understanding of the problem shows that the
contrast in the energy density freezes.\footnote{Actually all these
  statements are gauge dependent.  We can always choose to work in
  e.g. the uniform density gauge, where the contrast in the energy
  density vanishes identically at all scales, or in e.g.  the
  synchronous gauge, where the contrast in the energy density grows
  even for those modes larger than the Hubble radius.  It is only in
  terms of the conformal-Newtonian gauge that the behavior outlined in
  the previous paragraph makes sense, and it is only in this gauge
  that we can easily compare our results with those sketched in the
  three points above.}

Perfect fluid structure formation then demands matter with a low speed
of sound in order to have a window, $c_s/H < \lambda < 1/H$, where the
Jeans mode is released and the perturbations can grow.  Cold dark
matter (CDM) represents the simplest realization of this scenario, for
which one assumes $c_s^2\approx 0$ for all relevant modes.\footnote{If
  DM consists of collisionless particles it is more appropriate to
  talk about a free-streaming, rather than a Jeans, length; however,
  the idea is similar, see e.g. Sections~10.2 and~10.3 in
  Ref.~\cite{Coles} for a discussion.}

For a canonical scalar field we have $c_s^2=1$, see
e.g. Ref.~\cite{Garriga}, and it is usually argued that this implies
that small perturbations in the energy density do not
grow~\cite{not-grow}. This would seem to be a very serious argument
against the whole scalar field DM program~\cite{SFDM}, so we find it
mandatory to clarify the issue.

\begin{figure}[t]
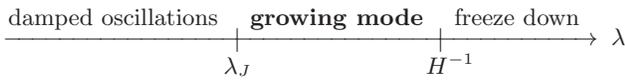

damped oscillations \hspace{.2cm} {\bf growing mode} \hspace{.2cm} freeze down $\quad$
\vspace{-4mm}
\[ \xrightarrow{\hspace*{7.7cm}} \; \lambda \] \\
\vspace{-6.5mm} $\qquad\quad |$ \hspace{2.4cm} $|\quad$ \\
\hspace{0.9cm} $\lambda_J$ \hspace{2.1cm} $H^{-1}\;$
\caption{Behavior of the linear perturbations in the contrast to the
  energy density for different wavelengths $\lambda$.  The Jeans
  length $\lambda_J$ depends on the matter content in the universe.
  For a perfect fluid $\lambda_J\sim c_sH^{-1}$, and structure
  formation demands $c_s^2\ll 1$.  However, when a canonical scalar
  field oscillates with high frequency around the minimum of the
  potential we have $\lambda_J\sim m^{-1}$, even though $c_s^2=1$. As
  usual this picture should be understood in the conformal-Newtonian
  gauge.}
\label{Fourier}
\end{figure}

In this paper we will consider the simplest situation: a universe
dominated by a real massive scalar field $\varphi$ satisfying the
Klein-Gordon equation, $(\Box+m^2)\varphi=0$.  Here the box is the
d'Alembert operator in four spacetime dimensions, and $m$ is the mass of
the scalar particle [this corresponds to a scalar field potential of
the form $V(\varphi)=m^2\varphi^2/2$, so that the mass is defined as
$m^2\equiv \partial^2 V/\partial\varphi^2$].  The more interesting
case with a complex field including the self-interactions 
and the presence of other matter components (e.g. radiation, baryons, etc) will be
presented elsewhere.  We will show that when the scalar field is
slowly rolling down the potential, linear perturbations cannot grow,
in accordance with common wisdom.  However, when the scalar field
oscillates with high frequency $m\gg H$ around the minimum of the
potential term, the Jeans length is not given by the naive value $c_s
H^{-1}$ one would guess from a perfect fluid analogy. Instead it is
determined by the Compton wavelength of the scalar particle, $c_s
m^{-1}$, and the evolution of perturbations larger than this scale
almost mimics that of the standard CDM scenario even though $c_s^2=1$ (see
Figure~\ref{Fourier} for details). The
  main difference with respect to the standard CDM evolution (apart
  from the appearance of a nonvanishing Jeans length), is the presence
  of a time-dependent modulation of the different physical quantities.
We will present below both an analytical argument based on expansions
of the solution of the relevant cosmological equations, and results
from simple numerical simulations.

Similar results have been presented before in the literature, see for
instance Refs.~\cite{previous} (as far as we know the gravitational
instability of a canonical scalar field ---in the context of a static
universe--- was reported for the first time by Khlopov, Malomed and
Zel'dovich in Ref.~\cite{Zeldovich}). However, to our knowledge, we
present for the first time a description of the problem in terms of
gauge-invariant fields. Furthermore, note that our analysis does not
rely on time averages and/or fluid identifications. Instead we present
a well-defined series expansion that makes it possible to find
analytic solutions order by order in the expansion parameter: it is
well known that the time average of the product of two functions is
not in general the product of the time averages of those functions, and also that
a scalar field is not a perfect fluid~\cite{diez-tejedor}, so one needs
to take some care with the standard procedure. With this formalism the
results emerge more naturally than in previous works, and it is
convenient in order to identify the root of the misunderstanding; see
Eq.~(\ref{eq.v}) and the paragraphs below.

We will follow the notation in Chapters~7 and~8 of
Ref.~\cite{Mukhanov}; in particular we use the signature $(+,-,-,-)$
for the spacetime metric.  We highly recommend this reference to the
reader interested in the details about some of the definitions and conventions below.
One should also mention that recently the problem of structure
formation with a massive scalar field has been studied by performing
full nonlinear numerical simulations of the Einstein-Klein-Gordon
system, both in the relativistic~\cite{nosotros} and the
nonrelativistic~\cite{Tom} regimes.  However, we believe that by
studying the problem from the point of view of a mode analysis in
perturbation theory one can more clearly separate the relevant
physical mechanisms that come into play at different scales.


\section{The homogeneous and isotropic background}
\label{sec:background}

At very large scales the universe is (nearly) homogeneous and
isotropic; that makes it possible to introduce the idea of a
homogeneous and isotropic background.  According to the current 
cosmological observations this background can be
described in terms of a flat Robertson-Walker (RW) metric of the form
\begin{equation}\label{eq.linea.homogeneous}
ds^2 = a^2 \left( d\eta^2 - dx^2 - dy^2 - dz^2\right) \,.
\end{equation}
Here $\eta$ is the conformal cosmological time, $(x,y,z)$ a spatial
coordinate system comoving with the expansion, and $a(\eta)$ the scale
factor. Conformal time $\eta$ is related to the standard comoving
cosmological time $t$ through $t = \int a d \eta$. The expansion rate
is codified in the Hubble parameter, $H=\mathcal{H}/a=a'/a^2$, with
the prime denoting the derivative with respect to conformal time. The
spacetime symmetries in Eq.~(\ref{eq.linea.homogeneous}) guaranty that
the background field cannot depend on the spatial
coordinates, so that $\varphi(\eta,\vec{x})=\varphi_0(\eta)$.  Under
these assumptions the Klein-Gordon equation simplifies to
\begin{equation}\label{eq.KG.homogeneos}
\varphi''_0 + 2 \mathcal{H} \varphi'_0 + a^2 m^2 \varphi_0 = 0 \,.
\end{equation}

There are two different timescales in Eq.~(\ref{eq.KG.homogeneos}): on
the one hand that associated to the cosmological expansion,
$\mathcal{H}^{-1} = (aH)^{-1}$, and on the other that defined by the
mass of the scalar field, $(am)^{-1}$.  In this paper we will
concentrate on the case when the expansion rate of the universe is
very slow when compared to the characteristic time of oscillation of
the scalar field, that is $(aH)^{-1} \gg (am)^{-1}$, which implies $H
\ll m$. This inequality is always satisfied at late times. We will
briefly discuss the regime $H\gg m$ later on in this section.

In order to proceed we propose a solution of the form
\begin{subequations}\label{eq.ansatz}
\begin{eqnarray} 
\varphi_0(\eta) &=& \varphi_0^H(\eta) \left[ \varphi_0^m(\eta)
+ \mathcal{O}^2(H/m) \right] \, , \\
\mathcal{H}(\eta) &=& \mathcal{H}^H(\eta) \left[ 1
+ \mathcal{O}(H/m) \right] \, , \label{eq.ansatz.H} \\
a(\eta) &=& a^H(\eta)\left[1 + \mathcal{O}^2(H/m)\right]\, .
\label{eq.ansatz.a}
\end{eqnarray}
\end{subequations}
Here functions with a superscript $H$ vary on cosmological timescales,
$(f^H)'\sim a H f^H$, whereas those with a superscript $m$ vary on
timescales given by the mass of the scalar field, $(f^m)'\sim a mf^m$.
In principle, higher-order terms oscillate in time with high
frequency. Consequently, the derivative of one of those terms is not
necessarily suppressed in the series expansion,
e.g.~$d[\mathcal{O}(H/m)]/d\eta\sim aH$.  That is the reason for which
the series expansion of the Hubble parameter $\mathcal{H}$ has a
linear term in $H/m$, even though such a term is not present in the
expression for the scale factor $a$.

Introducing the ansatz~(\ref{eq.ansatz}) into the Klein-Gordon
equation we obtain
\begin{equation}\label{eq.varphi}
 \varphi_0(\eta) = \frac{AM_{\textrm{Pl}}}{(m\eta)^3} \left[ \sin(mt)
+ \mathcal{O}^2(H/m) \right] \, ,
\end{equation}
with $A$ an integration constant, $M_{\textrm{Pl}}=1/\sqrt{8\pi G}$
the reduced Planck mass, and where we have fixed an arbitrary phase to
zero.  For later convenience, and with no loss of generality, from now
on we will also choose $A^2=24$.  From the above expression we see
that the scalar field oscillates around $\varphi_0=0$ with constant
frequency $m$ in comoving time $t$ (so that the frequency will
increase in conformal time $\eta$ as the universe expands), and an
amplitude that decays as $1/\eta^3$. Introducing this expression for
$\varphi_0(\eta) $ into the Friedmann equations,
\begin{subequations}\label{eq.friedmann}
\begin{eqnarray}
\mathcal{H}^2 &=& \frac{8\pi G}{3} \: a^2\varepsilon_0 \, ,
\label{eq.friedmann1} \\
\mathcal{H}'- \mathcal{H}^2 &=& - 4\pi G a^2 ( \varepsilon_0+p_0 ) \, ,
\label{eq.friedmann2}
\end{eqnarray}
\end{subequations}
one finds
\begin{eqnarray}
\mathcal{H}(\eta) &=& \frac{2}{\eta}\left[ 1
- \frac{3}{4} \left( \frac{H}{m} \right) \sin(2mt)
+ \mathcal{O}^2(H/m) \right] , \label{eq.H} \\  
a(\eta) &=& (m\eta)^{2}\left[1 + \mathcal{O}^2(H/m)\right] .
\label{eq.a}
\end{eqnarray}
For completeness we can also integrate the expression for the comoving
time to obtain
\begin{equation}
mt = \int ma\, d\eta = \frac{(m\eta)^3}{3} \left[ 1
+ \mathcal{O}^3(H/m) \right] \, .
\label{eq:time}
\end{equation}

In the above equations we have used the fact that
the background energy density and pressure are given by
\begin{equation}
\varepsilon_0 = \frac{1}{2} \left( \frac{\varphi'^2_0}{a^2}
+ m^2 \varphi_0^2 \right) \, ,\quad p_0 = \frac{1}{2}
\left( \frac{\varphi'^2_0}{a^2} - m^2 \varphi_0^2 \right) .
\label{eq:energy}
\end{equation}
In particular we find that, to lowest order in the series expansion,
the background energy density redshifts with the inverse of the
comoving volume, \mbox{$\varepsilon_0 \sim 1/a^3$}, whereas for the
background pressure we obtain \mbox{$p_0 \sim (1/a^3)\cos(2mt)$}.
Note that to this same order we cannot distinguish the expansion rate
from that in a CDM universe, $\mathcal{H}=2/\eta$, even tough
$|p_0|\sim \varepsilon_0$ during the evolution.  Interestingly, this
does not depend on the mass of the scalar particle, as long as it is
large enough when compared to the expansion rate of the universe.

Since $H/m \sim 2 /(m\eta)^3$, the condition $H\ll m$ demands
$m\eta\gg 1$.  In terms of the scalar field, Eq.~(\ref{eq.varphi})
above, we obtain $|\varphi_0|\ll M_{\textrm{Pl}}$.  That guarantees
large values
\mbox{$|\epsilon_{\textrm{sr}}|=|\eta_{\textrm{sr}}|=2M_{\textrm{Pl}}^2/\varphi_0^2\gg
  1$} of the slow-roll parameters~\cite{Liddle}, with
$\epsilon_{\textrm{sr}}\equiv
(M_{\textrm{Pl}}^2/2)(\partial_{\varphi}V/V)^2$ and
\mbox{$\eta_{\textrm{sr}}\equiv
  M_{\textrm{Pl}}^2(\partial^2_{\varphi}V/V)$}, so we can safely
conclude that the universe is not in a period of slow-roll inflation,
as was evident from Eq.~(\ref{eq.H}).

At this point one should mention that the dominant terms in the
solution of the Friedmann-Klein-Gordon system given by
Eqs.~\eqref{eq.varphi}, \eqref{eq.H} and~\eqref{eq.a} coincide with
those of the exact solution found in Ref.~\cite{axel} for a scalar field
evolving in a universe dominated by a barotropic fluid such that
$H=2/(3\gamma t)$, with $\gamma$ constant (when $\gamma=1$ this background barotropic
fluid can be associated with the average energy density and pressure
of the scalar field itself). Something similar happens with the
solution reported in Ref.~\cite{turner}, obtained in terms of time
averages.  The main difference with our results is the fact that here
we have made explicit the existence of higher-order terms in $H/m$,
and in particular we have shown the first oscillating subdominant
contribution to the Hubble parameter.  Note however that this is not
only a purely academic question: the inclusion of the higher-order
terms will be crucial next for the understanding of the evolution of
the small perturbations.

\begin{figure}[t]
  \centering
  \includegraphics[width=.49\textwidth]{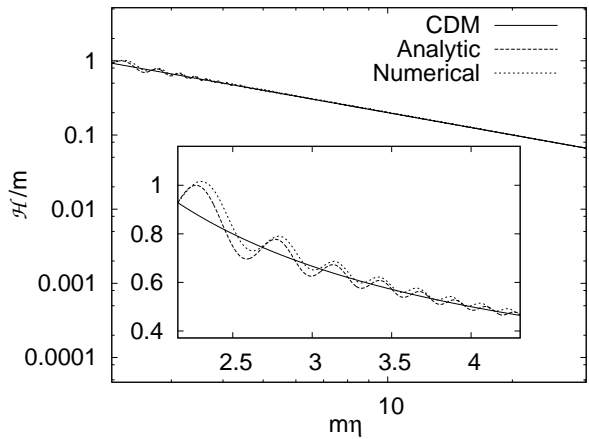}
  \caption{Evolution of the conformal Hubble factor $\mathcal{H}/m$ as
    a function of the conformal time $m\eta$ for a free scalar field
    oscillating around the minimum of the potential.  We show for
    comparison the solution in standard CDM (solid line), and both the
    analytical approximation in Eq.~(\ref{eq.H}) (dashed line), and
    the full numerical solution of the Friedmann-Klein-Gordon system,
    Eqs.~\eqref{eq.KG.homogeneos} and~\eqref{eq.friedmann}, (dotted
    line).  In the figure the integration begins at $m\eta=10^{1/3}$,
    with $a=10^{2/3}$ and $\mathcal{H}/m=2\times10^{-1/3}$.}
  \label{Hubble}
\end{figure}

Figure~\ref{Hubble} shows the evolution of the conformal Hubble factor
$\mathcal{H}/m$ as a function of the conformal time $m\eta$ for a free
scalar field oscillating around the minimum of the potential.  The
solid line corresponds to standard CDM, while the dashed and dotted
lines show the case of the scalar field using both the analytical
approximation in Eq.~(\ref{eq.H}), and a full numerical solution of
the Klein-Gordon and Friedmann system, Eqs.~\eqref{eq.KG.homogeneos}
and~\eqref{eq.friedmann}, respectively [for the numerical case we
solve for $\mathcal{H}$ from Eq.~\eqref{eq.friedmann2}, and use
Eq.~\eqref{eq.friedmann1} to monitor the numerical error].  Notice
that the evolution for the case of the scalar field follows closely
that of CDM with small oscillations around it. Initially we have
$H/m=0.2$ and the oscillations are still evident, but their amplitude
decreases as time goes on.

Next we will consider the behavior of the small perturbations around
the homogeneous and isotropic solution in Eqs.~(\ref{eq.varphi}),
(\ref{eq.H}), and~(\ref{eq.a}).  In particular, we find that an
unstable Jeans mode grows at scales between the Compton length of the
scalar particle and the Hubble radius of the universe; see
Eq.~(\ref{exp.densidad}) below for details.


\section{The linear perturbations}
\label{sec:perturbations}

In this paper we will not consider vector and tensor modes, and will
concentrate on the scalar sector of the perturbations. After all this
is the sector that contains the Jeans mode of the theory.  With this
assumption, the most general expression for the spacetime metric of a
universe close to a flat homogeneous and isotropic RW one takes the
form
\begin{eqnarray}\label{eq.line.perturbed}
ds^2 &=& a^2\left\lbrace(1+2\phi)d\eta^2
+ 2\partial_i B \: dx^i d\eta \right. \nonumber \\
&-& \left. \left[ (1-2\psi) \delta_{ij}
- 2\partial_i\partial_j E \right] dx^i dx^j \right\rbrace \,.
\end{eqnarray}
For the perturbed scalar field we will also write
\mbox{$\varphi=\varphi_0+\delta\varphi$}. Here $\phi$, $B$, $\psi$,
$E$ and $\delta\varphi$ are functions of the spacetime coordinates
$\eta$ and $\vec{x}$, with $\delta\varphi \ll \varphi_0$ and $\phi,\,
B,\, \psi,\, E \ll 1$.

All of these fields depend on the choice of coordinates used to write
the line-element in Eq.~(\ref{eq.line.perturbed}), and they could in
principle be describing fictitious inhomogeneities.  For this reason
it is sometimes convenient to work with gauge-invariant fields, such
as~\cite{Mukhanov}
\begin{subequations}\label{eq.gauge.fields}
\begin{eqnarray}
 \Phi &=& \phi-(1/a)\left[a(B-E')\right]' \,, \\
 \Psi &=& \psi + \mathcal{H}(B-E') \,, \\
 \overline{\delta\varphi} &=& \delta\varphi-\varphi_0'(B-E') \,.
\end{eqnarray}
\end{subequations}
Note that the above fields coincide with the amplitude of the metric
and the scalar field perturbations in the conformal-Newtonian (also
known in the literature as the longitudinal) coordinate system, for
which \mbox{$B=E=0$}.

To linear order in the new field variables, the $00$ and $0i$ Einstein
field equations, essentially the Hamiltonian and momentum constraints,
take the following form:
\begin{subequations}
\begin{eqnarray}
&& \Delta\Psi - 3 \mathcal{H} \left( \Psi' + \mathcal{H} \Phi \right)
= 4 \pi G a^2 \overline{\delta\varepsilon} \nonumber \\
&& \hspace{10mm} = 4 \pi G \left[ \varphi_0' \overline{\delta\varphi}\,'
+ a^2m^2 \varphi_0 \overline{\delta\varphi}
- \varphi_0'^2 \Phi \right] \, , \hspace{8mm} \label{eq:ham} \\
&& \Psi'+\mathcal{H}\Phi = 4\pi G \varphi_0'\overline{\delta\varphi} \, ,
\label{eq:mom}
\end{eqnarray}
\end{subequations}
where $\Delta$ is the Laplace operator in flat space, and where
$\overline{\delta\varepsilon}=\delta\varepsilon-\varepsilon_0'(B-E')$
is the gauge-invariant density perturbation. For a scalar field, and
to this order in the series expansion, there are no anisotropic
stresses and the $i\neq j$ field equations fix $\Phi=\Psi$. (Note
that so far we are using two different series expansions: one in the
small perturbations of the spacetime metric and the scalar field, and
the other in the expansion rate of the universe.  We will soon
introduce a new one in terms of the Jeans length.)

In order to move forward we find it convenient to define the new
quantities
\begin{equation}\label{eq.def.v,z}
v = a \left( \overline{\delta\varphi} + \frac{\varphi_0'}{\mathcal{H}}
\: \Psi \right) \, , \quad
z = \frac{a\varphi_0'}{\mathcal{H}} \, .
\end{equation}
The gauge-invariant field $v(\eta,\vec{x})$ is usually known as the
Mukhanov-Sasaki variable, and the function $z$ depends only on the
spacetime background, $z=z(\eta)$. Note that, leaving the scale factor
aside, the Mukhanov-Sasaki variable represents the scalar field
perturbation evaluated in the spatially flat gauge
$\psi=E=0$~\cite{Wands}.  Whenever the scalar field dominates the
evolution of the universe, Eqs.~\eqref{eq:ham} and~\eqref{eq:mom}
simplify to
\begin{subequations}
\begin{eqnarray}
\Delta \left( \frac{a^2\Psi}{\mathcal{H}} \right) &=&
4 \pi G z^2 \left( \frac{v}{z} \right)' \, ,
\label{eq:ham2} \\
\left( \frac{a^2\Psi}{\mathcal{H}} \right)' &=&
4 \pi G z^2 \left( \frac{v}{z} \right)  \, .
\label{eq:mom2}
\end{eqnarray}
\end{subequations}
These two equations can be combined to obtain
\begin{equation}\label{eq.v}
 v'' - c_s^2 \Delta v - \frac{z''}{z} \: v = 0 \; ,
\end{equation}
with $c_s^2=1$ analogous to a ``speed of sound'' for a canonical
scalar field~\cite{Garriga}. The main reason for this identification
is the similarity of Eq.~(\ref{eq.v}) with that obtained for a
barotropic perfect fluid with sound speed $c_s$; see e.g. Eq.~(7.65)
in Ref.~\cite{Mukhanov}.  Notice that even though here we are only
interested in the case of a scalar field with no self-interactions, so
far the analysis of the perturbations is general and valid for an
arbitrary potential.
 
Equation~(\ref{eq.v}) above fixes a characteristic length scale given
by \mbox{$c_s a|z''/z|^{-1/2}$} (in this paper we will work with
comoving wavenumbers $k$, but we will talk about physical wavelengths
$\lambda$, with $\lambda=2\pi a/k$).  This length provides us with a
ruler to discriminate between large and small scales in the scalar
field perturbations. The correct estimation of this length scale for
the different physical situations ---and in particular the relative
size it takes when compared to the Hubble radius--- lies precisely at
the heart of the usual misunderstanding.

As we will find soon, see Eqs.~(\ref{exp.psi.1})
and~(\ref{exp.densidad}) below, the scale $c_s a|z''/z|^{-1/2}$
determines the Jeans length of the scalar field. However, this
quantity is not always related to the naive guess $c_sH^{-1}$ one
would expect from a perfect fluid analogy. In order to clarify this
point, let us consider two representative cases.  When the scalar
field is rolling down to the minimum of the potential we have
$\varphi'_0\sim aH\varphi_0$, which implies $|z''/z|\sim (aH)^{2}$.
This fixes the characteristic Jeans length to the Hubble
radius. Short-wavelength perturbations (when compared to this length
scale) oscillate in space and time as \mbox{$v \sim \sin(c_s k \eta +
  \vec{k}\cdot \vec{x})$}, with $c_s^2=1$.  The Jeans mode is then
stabilized and the perfect fluid analogy seems possible, i.e. small
perturbations to the homogeneous and isotropic solutions do not grow.
Consider, for instance, the case of the inflaton during a slow-roll
regime, or a quintessence field in the present universe, as particular
realizations of this scenario.


However, when the zero-mode of the scalar field is oscillating with
high frequency $m\gg H$ around the minimum of the potential, we have
instead $\varphi'_0\sim am\varphi_0$, which results in $|z''/z| \sim
(am)^2$. The new length scale is well inside the Hubble radius, making
possible the growth of perturbations with \mbox{$aH < k < c_s^{-1}
  am$} (remember that for a canonical scalar field $c_s^2=1$).  Now
the naive estimation $c_sH^{-1}$ for the Jeans length has nothing to
do with the correct one, $c_sm^{-1}$.

Let us consider in more detail this second scenario.  Notice that, using
the solutions in Eqs.~(\ref{eq.varphi}),~(\ref{eq.H}),
and~(\ref{eq.a}), the ``mass'' term in Eq.~(\ref{eq.v}) takes the form
\begin{equation}
\frac{z''}{z} = - a^2 m^2 \left[ 1 + 6 \left(\frac{H}{m}\right) \sin(2mt)
+ \mathcal{O}^2(H/m) \right] .
\label{eq:z-sol}
\end{equation}
As mentioned before, this is a purely background quantity.
For practical reasons we will work with periodic boundary conditions
over a box of comoving size $L$.  We can always take the limit
$L\to\infty$ at the end of the calculations.  Under this assumption,
the general solution to Eq.~(\ref{eq.v}) can be written in the
form
\begin{equation}\label{expansion.v}
v(\eta,\vec{x}) = \frac{1}{L^{3/2}}\sum_{\vec{k}\neq 0}
\left[ C_{\vec{k}} v_{\vec{k}}(\eta) e^{i\vec{k}\cdot\vec{x}} + c.c. \right] \, .
\end{equation}
Here $C_{\vec{k}}$ are some dimensionless integration constants that
label the different possible solutions, $c.c.$ denotes complex
conjugate, and the functions $v_{\vec{k}}(\eta)$ satisfy the equation
\begin{equation}
\label{eq.v_k}
 v''_{\vec{k}} + \omega_{\vec{k}}^2(\eta) v_{\vec{k}} = 0 \, , \quad
\textrm{with} \quad \omega_{\vec{k}}^2(\eta) = k^2 - \frac{z''}{z} \, .
\end{equation}
Note that the boundary conditions demand \mbox{$k_i=2\pi n_i/L$}, with
$n_i = \pm 1,\pm 2,\ldots$ and $i=1,2,3$.  The zero-mode \mbox{$k=0$}
is already included in the description of the spacetime background,
and for that reason it does not appear in Eq.~(\ref{expansion.v}).

According to the Eq.~(\ref{eq.v_k}) the behavior of the mode-functions
$v_{\vec{k}}(\eta)$ depends crucially on the relative value between
the square of the wavenumber $k^2$, and the background function
$z''/z$.  For those modes smaller than the Compton wavelength, $k \gg
am$, we obtain (remember that in the fast expansion regime 
the Compton wavelength is always well inside the Hubble radius, and then $am \gg aH$)
\begin{widetext}
\begin{subequations} \label{exp.v}
\begin{equation} \label{exp.v.1}
 v_{\vec{k}}(\eta) = \frac{1}{\sqrt{2k}}\left( 1 + \mathcal{O}^2(H/m, am/k) \right)
\exp \left[ -i \left( 1 + \mathcal{O}^2(H/m, am/k) \right) k\eta \right] \, ,
\end{equation}
whereas for modes larger than this quantity, $k \ll am$, we find
\begin{eqnarray}\label{exp.v.2}
v_{\vec{k}}(\eta) &=& \frac{A(k)}{\sqrt{2m}} \left \lbrace B(k)\bar{z}
- \frac{1}{5}B(k)a \left( \frac{k}{am} \right) \left( \frac{k}{aH} \right)
\left[ \sin(mt) - \frac{3}{2} \left( \frac{H}{m} \right) \cos^3(mt)
+ \mathcal{O}^2(H/m) \right] \right. \nonumber \\
&& - iB^{-1}(k) \bar{z} \int\frac{d(m\eta)}{\bar{z}^2}
+ \frac{i}{7}B^{-1}(k)\frac{1}{a^2} \left( \frac{k}{am} \right) \left( \frac{k}{aH} \right)
\left[ \cos(mt) - \frac{3}{2} \left( \frac{H}{m} \right)
\sin^3(mt) + \mathcal{O}^2(H/m)\right] \nonumber \\ 
&& \left. + \mathcal{O}^2(H/m, k/am) \rule{0mm}{5mm} \right\rbrace \, ,
\end{eqnarray}
\end{subequations}
\end{widetext}
(or $v_{\vec{k}}=v^{\textrm{r}}_{\vec{k}}+iv^{\textrm{i}}_{\vec{k}}$
if we prefer to think in terms of two real, linearly independent
mode-functions $v^{\textrm{r}}_{\vec{k}}$ and
$v^{\textrm{i}}_{\vec{k}}$.)  Here we have defined
$\bar{z}=z/\sqrt{6}M_{\textrm{Pl}}$. Also, the functions
$v_{\vec{k}}(\eta)$ have been normalized so that
$v_{\vec{k}}(v_{\vec{k}}^*)'-v'_{\vec{k}}(v_{\vec{k}}^*)=i$, with the
convention that in the asymptotic limit $H/m\to 0$, $k/(am)\to\infty$,
we recover standard massless plane waves,
i.e. $v_{\vec{k}}(\eta)=e^{-ik\eta}/\sqrt{2k}$ in
Eq.~(\ref{exp.v.1}). In the above expression $A(k)$ is a phase factor,
$|A(k)|^2=1$, and $B(k)$ is a real number necessary to connect the two
regimes in Eqs.~(\ref{exp.v.1}) and~(\ref{exp.v.2}) at $k\sim am$. For
our purposes in this paper we will not need to determine these
quantities.

Notice that, while factors of $aH/k$ are small in Eq.~(\ref{exp.v.1}),
this is not necessarily true for factors of $k/aH$ in the case of
Eq.~(\ref{exp.v.2}), where they can even dominate for modes inside the
Hubble radius, i.e. those modes with \mbox{$aH < k < am$}. This
implies that, for modes larger than the Compton wavelength, the lowest
nonvanishing contribution to the Mukhanov-Sasaki variable is not
given by \mbox{$\lim_{k\to 0}v_{\vec{k}}(\eta)\sim\bar{z}-i\bar{z}\int
  d(m\eta)/\bar{z}^2$}, as one could have naively expected.
Incidentally the extra terms will be crucial to determine the behavior
of the modes relevant for structure formation. Since $aH \sim
\eta^{-1}$ and $am \sim \eta^2$, eventually all modes reach this
regime. This completes our discussion of the Mukhanov-Sasaki
variable.

However, the Mukhanov-Sasaki variable is only an auxiliary field
(remember that this quantity is related to the perturbation in the
scalar field evaluated in the spatially flat gauge).  In order to make
contact with observations we need to move our attention, for instance,
to the Newtonian potential $\Psi$, or to the contrast in the energy
density, $\overline{\delta\varepsilon}/\varepsilon_0$.  Introducing
the expressions for the Mukhanov-Sasaki variable, Eqs.~(\ref{exp.v}),
into e.g.  Eq.~(\ref{eq:ham2}), we obtain for the Newtonian
potential\footnotemark[4]
\begin{widetext}
\begin{equation}\label{exp.psi.1}
\Psi_{\vec{k}}(\eta) = \sqrt{\frac{3}{2}} \: \frac{1}{M_{\textrm{Pl}}} \times
\left\{ \begin{array}{ll}
\displaystyle{ \frac{i}{\sqrt{2k}} \left( \frac{aH}{k} \right)
\frac{1}{a} \: \cos(mt) \: e^{-ik\eta}} \: ,
& \mbox{if $k\gg am$} \, , \\
\displaystyle{ \frac{A(k)}{\sqrt{2m}} \left[\frac{1}{5} B(k) + iB^{-1}(k) \left( \frac{aH}{k} \right)
\left( \frac{am}{k} \right) \frac{1}{a^3} \right]} \: ,
& \mbox{if $k\ll am$} \, .
\end{array} \right.
\end{equation}

\noindent whereas from the Hamiltonian constraint~(\ref{eq:ham}), we find for
the contrast in the energy density\footnotemark[5]
\begin{equation}\label{exp.densidad}
\left.\frac{\overline{\delta \varepsilon}}{\varepsilon_0}\right|_{\vec{k}}
= \sqrt{\frac{2}{3}} \frac{1}{M_{\textrm{Pl}}} \times
\left\{ \begin{array}{ll}
\displaystyle{ \frac{-i}{\sqrt{2k}} \left( \frac{k}{aH} \right)
\frac{1}{a} \: \cos(mt) \: e^{-ik\eta}} \: , & \mbox{if $k \gg am$} \, , \\
\displaystyle{ \frac{A(k)}{\sqrt{2m}} \left[ - \frac{1}{5}B(k) \left( \frac{k}{aH} \right)^2
- iB^{-1}(k) \left( \frac{m}{H} \right) \frac{1}{a^3} \right]} \:, & \mbox{if $aH \ll k \ll am$} \, , \\
\displaystyle{ \frac{A(k)}{\sqrt{2m}} \left[ -\frac{6}{5}B(k) + 9 iB^{-1}(k) \left( \frac{aH}{k} \right)
\left( \frac{am}{k} \right) \frac{1}{a^3} \right] \cos^2(mt)} \: , & \mbox{if $k\ll aH$} \, .
\end{array} \right.
\end{equation}
\end{widetext}
Note that the two expressions above have been reported only to the
lowest non-vanishing order in the series expansions.
A pattern of small-amplitude, high-frequency oscillations are expected around
the expressions in Eqs.~(\ref{exp.psi.1}) and~(\ref{exp.densidad}), but this will be
enough for the purposes of this paper.

\footnotetext[4]{Here we have used the identity
\begin{equation}\label{exp.psi.general}
\Psi_{\vec{k}}(\eta) = -8\pi G\: \frac{z^2}{a^2} \left[ \frac{\mathcal{H}}{2k^2}
\left(\frac{v_{\vec{k}}}{z}\right)'\right] \,,
\end{equation}
that relates, in Fourier space, the Mukhanov-Sasaki variable to the
Newtonian potential.}

For modes larger than the Compton wavelength, \mbox{$k < am$}, the
Newtonian potential mimics the behavior of standard CDM: one of the
solutions remains constant, while the other decreases in time with
$H/a\sim\eta^{-5}$; see for instance Eq.~(7.53) in
Ref.~\cite{Mukhanov} for details.  On the other hand, for modes
smaller than the Compton wavelength, $k > am$, the solution oscillates
in time with a decreasing amplitude, $H\sim 1/\eta^3$. This decay in
the amplitude of the Newtonian potential is characteristic of a
barotropic perfect fluid $p=p(\varepsilon)\ll \varepsilon$ with a
nonvanishing Jeans scale, $c_s^2\neq0$.

Something similar happens for the contrast in the energy density,
where high-frequency modes with $k > am$ oscillate in time with damped
amplitude, \mbox{$1/a^2H\sim \eta^{-1}$}. In contrast, there appear
two different regimes for the modes larger than the Compton
wavelength. Perturbations larger than the Compton length but still
smaller than the Hubble radius, $aH < k < am$, can grow in time as
$1/(aH)^2 \sim \eta^2$, or decrease as $1/(Ha^3)\sim \eta^{-3}$, which
is again the same behavior as in the standard CDM scenario.  On the
other hand, those modes larger than the Hubble radius, $k < aH$,
freeze with a time-dependent modulation in $\cos^2(mt)$.  It is
possible to trace back this oscillatory dependency to a background
term, \mbox{$\mathcal{H}'-\mathcal{H}^2\sim \cos^2(mt)/a$}, that
appears in the expression for the contrast in the energy density, see
Eqs.~(\ref{eq.contraste.general}) and~(\ref{eq.grandes-pequenas}) in
footnote~\ref{note:contraste}. These oscillations are therefore not
expected to appear in the case of a complex scalar field for which
\mbox{$\mathcal{H}'-\mathcal{H}^2\sim 1/a$}.

\footnotetext[5]{\label{note:contraste}
Here we have used the identity
\begin{equation}
\label{eq.contraste.general}
\left.\frac{\overline{\delta \varepsilon}}{\varepsilon_0}\right|_{\vec{k}}
= 8\pi G\:\frac{z^2}{a^2} \left[\left(\frac{1}{3}
+ \frac{\mathcal{H}' - \mathcal{H}^2}{k^2}\right)
\frac{1}{\mathcal{H}}\left(\frac{v_{\vec{k}}}{z}\right)' 
- \left(\frac{v_{\vec{k}}}{z}\right)\right] ,
\end{equation}
that relates, in Fourier space, the Mukhanov-Sasaki variable to the
contrast in the energy density.  Using the evolution for the
background universe we can write
\begin{equation}\label{eq.grandes-pequenas}
\frac{1}{3} + \frac{\mathcal{H}'-\mathcal{H}^2}{k^2} = 
\frac{1}{3}\left[1-9\left(\frac{aH}{k}\right)^2\cos^2(mt)
+ \mathcal{O}(H/m)\right] .
\end{equation}}

In particular, the growing mode in the energy density contrast
$\overline{\delta \varepsilon}/\varepsilon_0 \sim 1/(aH)^ 2 \sim
\eta^2 \sim t^{2/3}$ (i.e. \mbox{$\overline{\delta
    \varepsilon}/\varepsilon_0\sim a$} to the lowest order in the
expansion series) at those scales larger than the Compton wavelength
but smaller than the Hubble radius, $1/m< \lambda< 1/H$, is usually
identified with structure formation in the universe (see for instance
Eq.~(7.56) in Ref.~\cite{Mukhanov}).  Note however that the (large)
oscillations for the contrast in the energy density of modes larger
than the Hubble radius are not present in the case of CDM.

\begin{figure}[t]
\centering
\includegraphics[width=.49\textwidth]{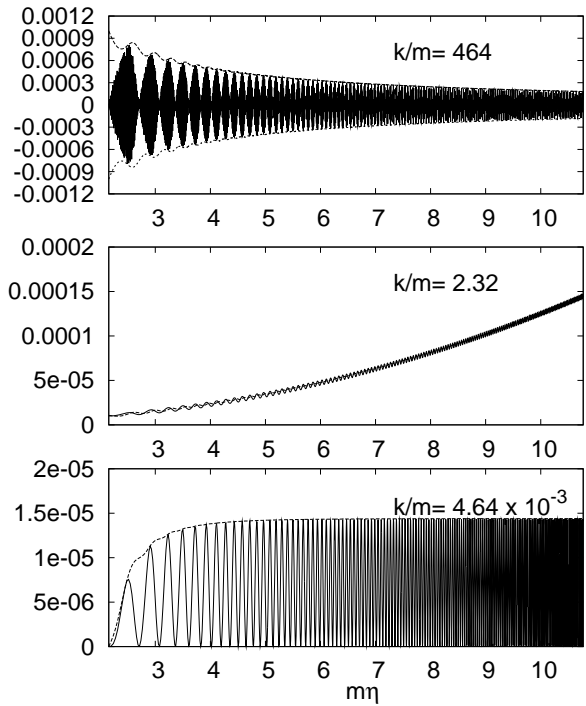}
\caption{Evolution of the contrast in the energy density,
  $\overline{\delta \varepsilon}/\varepsilon_0|_{\vec{k}}$, as a
  function of conformal time $m\eta$ for three different Fourier
  modes, $k/m=(10^2,\,1/2,\,10^{-3})\times 10^{2/3}$, in the universe
  depicted in Figure~\ref{Hubble}. The lower and higher wavenumbers
  were chosen in order to lie outside of the region $[aH,am]$ during
  the evolution.  Solid lines represent the results of the numerical
  evolution, whereas dashed lines are obtained from an appropriate
  combination of the envelopes of the linearly independent solutions
  in Eq.~(\ref{exp.densidad}).  Modes shorter than the Compton
  wavelength (top panel) oscillate with damped amplitude.  On the
  contrary, modes larger than the cosmological horizon (bottom panel)
  freeze, showing an oscillatory behavior associated to the inherent
  oscillation of the scalar field. Finally, those modes that are
  larger than the Compton wavelength but smaller than the Hubble
  radius (middle panel) grow at the same rate as the scale factor, as
  it is expected in the case of CDM. These modes can give rise to
  structures in the late universe. Note that since the equations are
  linear we can rescale the vertical axes in the figures arbitrarily,
  as long as the amplitudes remain always less than unity.}
\label{k.modes}
\end{figure}

At this point it is interesting to stress once again the fact that these
statements are only valid in the conformal-Newtonian coordinate
system, where the gauge-invariant fields $\Psi$ and
$\overline{\delta\varepsilon}/\varepsilon_0$ take the same values as
the Newtonian potential and contrast in the energy density,
respectively.  If we move to e.g. the synchronous coordinate system
$\phi=B=0$, the contrast in the energy density is related to the
gauge-invariant field $\overline{\delta\varepsilon}/\varepsilon_0$
through
$\delta\varepsilon_{\textrm{s}}/\varepsilon_0=\overline{\delta\varepsilon}/\varepsilon_0-
\varepsilon_0'/(a\varepsilon_0)\int a\Psi d\eta$, the constant of
integration in this formula corresponding to an unphysical fictitious
mode.  As in the standard CDM scenario, the expression for
intermediate wavelengths $aH \ll k \ll am$ in Eq.~(\ref{exp.densidad})
describes now all the modes larger than the Compton wavelength of the
scalar particle, $k \ll am$, and then in this gauge the contrast in
the energy density grows with time as the scale factor, even for those
modes that are well outside the Hubble horizon. (It is in this gauge
that the matter power spectrum is usually presented in the literature.)
Note that in this coordinate system the large oscillations in the
energy density disappear, and then it is not clear for us if they could be observable in
practice or if they are only a gauge artifact.

Figure~\ref{k.modes} shows the evolution of the contrast in the energy
density, $\overline{\delta \varepsilon}/\varepsilon_0|_{\vec{k}}$, for
three individual Fourier modes that are representative of the
different regimes. For this figure we have considered the same
background evolution of Figure~\ref{Hubble}, so that initially we have
\mbox{$m\eta=10^{1/3}$}, $a=10^{2/3}$ and $\mathcal{H}/m=2\times
10^{-1/3}$.  In order to calculate the contrast in the energy density
we integrate numerically Eqs.~\eqref{eq:mom2} and~\eqref{eq.v_k} for
the function $v_{\vec{k}}$ and the Newtonian potential
$\Psi_{\vec{k}}$, simultaneously with the background evolution.  We
then use the Hamiltonian constraint~\eqref{eq:ham} to find the
perturbation in the energy density, $\overline{\delta
  \varepsilon}_{\vec{k}}$.

The top panel in Figure~\ref{k.modes} corresponds to a mode with
\mbox{$k/m= 10^{8/3}\approx 464$}, which is clearly in the regime $k >
am$ for the time interval $m\eta\in (2,11)$. We can see that the
contrast in the energy density oscillates with an amplitude that
decays like $1/(a^2H)$, as expected. One can also clearly see from the
figure that we have oscillations with two quite different frequencies:
a high frequency oscillation coming from the term $e^{-ik \eta}$,
modulated by a lower frequency oscillation that corresponds to the
term $\cos(mt)$ in Eq.~(\ref{exp.densidad}). The middle panel
corresponds to a mode with $k/m=1/2\times10^{2/3}\approx 2.32$, so
that $aH < k < am$, also in the same time interval.  In this case we
see that the contrast in the energy density grows as $1/(a^2H^2)$,
with small oscillations.  These modes are interesting for structure
formation since they grow as if they were made of CDM.  Finally, the
bottom panel corresponds to a mode with $k/m= 10^{-7/3}\approx 4.64
\times 10^{-3}$, for which we have $k < aH$ in the evolution.  In this
case the contrast in the energy density rapidly reaches a regime where
it oscillates with constant amplitude.  Notice that in all these three
scenarios the frequency of the oscillations increases as time goes
by. This is again to be expected since the frequency should be
constant in cosmological time $t$, so that it increases in conformal
time~$\eta$.


\section{Discussion}

For a perfect fluid with equation of state
$p=p(\varepsilon)\ll\varepsilon$, there is a close relation between
the Hubble radius and the Jeans length, $\lambda_J \sim c_s H^{-1}$.
This is no longer true in the case of a canonical scalar field, where
the characteristic Jeans length is fixed instead by $\lambda_J \sim
c_s a|z''/z|^{-1/2}$.  When this length scale is much smaller than the
Hubble radius, perturbations can grow.  For a perfect fluid this is
possible only if $c_s^2 \ll 1$. For a scalar field we have $c_s^2 =1$,
but we could still satisfy $a|z''/z|^{-1/2}\ll H^{-1}$. This is what
happens, for instance, when the zero-mode of the scalar field is
oscillating with high frequency (when compared to the expansion rate
of the universe) around a minimum of the potential, where
$a|z''/z|^{-1/2} \sim m^{-1}$.  Note that this value for the Jeans
length at the scale of the Compton wavelength of the scalar particle
cannot be resolved in terms of a nonrelativistic analysis, as it was
previously done by Hu {\it et al.}  in Ref.~\cite{SFDM}.  Indeed, the
value of the Jeans length we identify in this paper does not coincide
with the naive estimation they reported in Eq.~(4) of that reference,
and which has been frequently used after that (see
e.g. Ref.~\cite{Tom}).

For those modes larger than the Jeans length the scalar field follows
the evolution in the standard CDM scenario, except for the large
oscillations of the contrast in the energy density for modes larger
than the Hubble radius when evaluated in e.g. the conformal-Newtonian
gauge. Note however that these (large) oscillations are not present in
the behavior of the Newtonian potential, for which the scalar field
behaves like CDM, or even in the super-Hubble modes of the contrast to
the energy density when evaluated in e.g. the synchronous gauge.  As
far as we know these oscillations of the contrast in the energy
density for long wavelength modes in certain coordinate systems had
not been previously reported in the literature, and it will be very
interesting to explore if they are only a gauge artifact or if on the
contrary they could have imprinted some signatures on the cosmological
observables. We expect these oscillations will not affect the large
scale structure of the universe. After all, the distribution of
galaxies is only sensitive to the subhorizon modes, and these modes
follow the standard CDM evolution.  However, they could affect
e.g. the cosmic microwave background photons at large angular scales,
which traced super-Hubble scales for a long period of time during the
cosmological evolution.  We leave a more detailed analysis of this
point for a future work.

For those modes smaller than the Jeans length the evolution cannot
bring the small perturbations in the early universe to the nonlinear
regime, and the inhomogeneities are erased.  This will introduce a
cutoff in the mass power spectrum for the distribution of galaxies in
the universe.  Something similar happens in warm DM
scenarios~\cite{wdm}.  If the mass of the scalar particle lies at the
scale of 10$\,\mu$eV or above, as it is expected for the QCD
axion~\cite{Gondolo}, the Jeans length would be smaller than a
centimeter, and the growth of cosmic structures would be probably
indistinguishable to that in the standard CDM scenario, at least while
in the linear regime (see Ref.~\cite{nosotros} for the case when
nonlinearities become important).  However, if we consider ultralight
scalar particles of masses as low as $10^{-22}\,$eV~\cite{SFDM,
  Arvanitaki:2009fg}, the Jeans length grows to the scale of parsecs.
This could have observable physical consequences in
cosmology~\cite{Tom, marsh}, alleviating, for instance, the missing satellite
discrepancy~\cite{satellites}.

In order to determine properly the new expression for the mass power
spectrum we would need a more elaborate analysis that includes, on
the one hand, a knowledge of the initial conditions for the scalar
field after inflation, as well as the evolution of the perturbations
during the radiation dominated era. We leave this study for a future
paper.

{\it Note added.}---After the first version of this paper was submitted
for publication we became aware of Ref.~\cite{Easther}, where the
linearized Einstein-Klein-Gordon system is considered in the context
of cosmological reheating. Although there are some similarities in the
analysis, the motivation of our paper is different.  We thanks Prof.~James
P. Zibin for pointing this reference out.


\begin{acknowledgments}
We are grateful to Juan Carlos Hidalgo, David
Marsh and Dar\'io N\'u\~nez for useful comments and discussions about a first version of
this paper.  This work was partially supported by PIFI, PROMEP,
DAIP-UG, CAIP-UG, the ``Instituto Avanzado de Cosmolog\'ia'' (IAC)
collaboration, DGAPA-UNAM under Grants No.~IN115311 and No.~IN103514,
CONACyT M\'exico under grants No. 182445 and No. 167335, and PAPIIT-IN101415.  ADT is
also supported in part by Grant No.~FQXi-1301 from the Foundational
Questions Institute (FQXi).
\end{acknowledgments}



\end{document}